\documentclass[12pt,preprint]{aastex}

\newcommand{\htwoo}{H$_2$O}
\newcommand{\htwoco}{H$_2$CO}

\newcommand{\ntwo}{N$_2$}

\newcommand{\cotwo}{\mbox{CO$_{2}$}}

\newcommand{\xsig}{$X\,^1\Sigma ^{+}$}

\newcommand{\api}{$A\,^1\Pi$}

\newcommand{\kms}{km~s$^{-1}$}

\newcommand{\mols}{molecules~s$^{-1}$}
\newcommand{\iue}{\textit{IUE\/}}
\newcommand{\hst}{\textit{HST\/}}

\shorttitle{CO in Comet 9P/Tempel~1}
\shortauthors{Feldman et al.}

\begin{document}

\title{Carbon Monoxide in Comet 9P/Tempel~1 before and after
the Deep Impact Encounter\altaffilmark{1}}

\author{Paul D. Feldman,\altaffilmark{2} Roxana E.
Lupu,\altaffilmark{2} Stephan R. McCandliss,\altaffilmark{2} Harold A.
Weaver,\altaffilmark{3} Michael F. A'Hearn,\altaffilmark{4} Michael J. S. Belton,\altaffilmark{5} and Karen J. Meech\altaffilmark{6}}


\altaffiltext{1}{Based on observations with the NASA/ESA {\it Hubble
Space Telescope} obtained at the Space Telescope Science Institute,
which is operated by the Association of Universities for Research in
Astronomy (AURA), Inc., under NASA contract NAS 5-26555.}

\altaffiltext{2}{Department of Physics and Astronomy, The Johns Hopkins
University, Charles and 34th Streets, Baltimore, MD 21218,
pdf@pha.jhu.edu, roxana@pha.jhu.edu, stephan@pha.jhu.edu}

\altaffiltext{3}{Space Department,
Johns Hopkins University Applied Physics \mbox{Laboratory,}
11100 Johns Hopkins Road,
Laurel, MD 20723-6099, hal.weaver@jhuapl.edu}

\altaffiltext{4}{Department of Astronomy, University of Maryland, 
College Park MD 20742-2421, ma@astro.umd.edu}

\altaffiltext{5}{Belton Space Exploration Initiatives,
Tucson, AZ 85716, mbelton@dakotacom.net}

\altaffiltext{6}{University of Hawaii, Institute for Astronomy and 
NASA Astrobiology Institute, 
2680 Woodlawn Drive, Honolulu, HI 96822, meech@ifa.hawaii.edu}



\begin{abstract}

One of the goals of the {\it Hubble Space Telescope} program to observe
periodic comet 9P/Tempel 1 in conjunction with NASA's Deep Impact
mission was to study the generation and evolution of the gaseous coma
resulting from the impact.  For this purpose, the Solar Blind Channel
of the Advanced Camera for Surveys was used with the F140LP
filter which is sensitive primarily to the ultraviolet emission ($\ge
1400$\AA) from the CO Fourth Positive system.  Following the impact we
detected an increase in brightness, which if all due to CO corresponds
to $1.5 \times 10^{31}$~molecules or a mass of $6.6 \times 10^5$ kg, an
amount that would normally be produced by 7--10 hours of quiescent
outgassing from the comet.  This number is $\leq$10\% of the number of
water molecules excavated, and suggests that the volatile content of
the material excavated by the impact did not differ significantly from
the surface or near sub-surface material responsible for the quiescent
outgassing of the comet.

\end{abstract}


\newpage
\section{INTRODUCTION}

The Deep Impact mission \citep{A'Hearn:2005a,A'Hearn:2005b}
successfully placed a 364~kg impactor, onto the surface of comet 9/P
Tempel 1 at a relative velocity of 10.3~\kms\ on 2005 July 4 at
05:52:03 UT (as seen from Earth).  The event was observed by cameras
aboard the mother spacecraft and by a large number of Earth- and
space-based telescopes as part of an extensive campaign to study the
comet prior to, during, and in the course of several days following the
impact \citep{Meech:2005}.  The {\it Hubble Space Telescope} (\hst)
provided the highest spatial resolution images (36 km) from Earth using
the Advanced Camera for Surveys (ACS) High Resolution Channel (HRC),
and also the possibility of using ultraviolet spectroscopy to study the
evolution of gaseous species, particularly CO, released by the impact.
The latter was initially planned for the Space Telescope Imaging
Spectrograph (STIS), but after the failure of STIS in August 2004, was
replaced with filter and prism observations using the Solar Blind
Channel (SBC) of ACS.  This paper presents the results of the SBC
observations and an estimate of the amount of CO released in the
gaseous ejecta.  A discussion of the visible imaging results is given
by \citet{Feldman:2006}.

\section{OBSERVATIONS}

The \hst\ campaign consisted of 17 separate ``visits'' during 2005
June and July, each comprising
a single \hst\ orbit allowing $\sim$53~minutes of target visibility per
orbit.  The July program (ID 10456) was divided into three periods: a
pre-impact group beginning roughly one cometary rotation before impact
to establish a baseline for the data to follow; the \hst\ orbit that
included the impact time; and several orbits immediately following the
impact and continuing, with single orbits, 7 and 12 days after the
impact.  Because of the large overhead in switching cameras during a
single \hst\ orbit, entire visits were dedicated to the SBC
observations.  These included two orbits prior to impact and three
within the 24 hour period following impact, which it turns out were
sufficient to see the ultraviolet brightness of the coma return to its
quiescent level.  The observations were made with both a long-pass
filter (F140LP) that is sensitive primarily to the ultraviolet emission
from the Fourth Positive system of carbon monoxide, and with the
largely unused objective prism PR130L.  The filter has a peak
throughput at 1390~\AA\ and half-power points at 1362 and 
1584~\AA.\footnote[7]{see the ACS Instrument Handbook available at \\ http://www.stsci.edu/hst/acs/documents/handbooks/cycle15/c10\_ImagingReference41.html\#317506 }
Because of the low count rate of the data, the prism data are
difficult to interpret, and in the following we will focus only on the
images obtained with the long-pass filter.  A log of the images is
given in Table~\ref{co_table}.
 

\section{ULTRAVIOLET IMAGING}

\subsection{Data Analysis}

\begin{figure}
\begin{center}
\epsscale{0.7}
\rotatebox{0.}{
\plotone{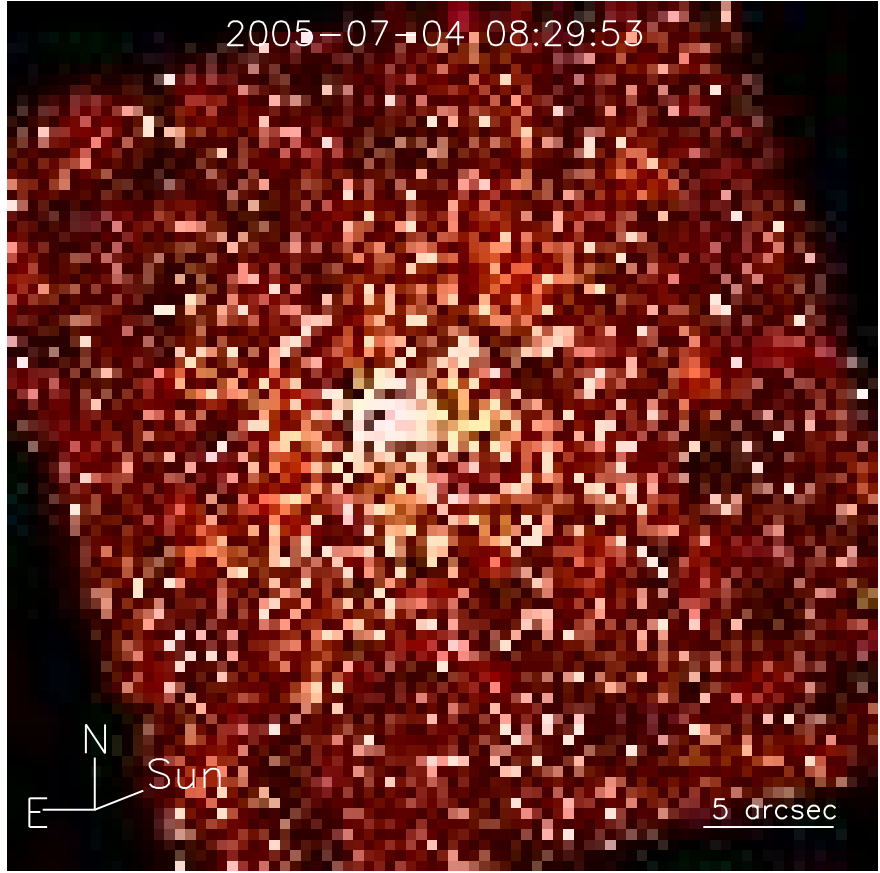}}
\caption{F140LP image of comet 9P/Tempel~1 taken 2.69~hr after the
impact.  The image has been geometrically corrected and rotated and
then rebinned to $0.\!''4 \times 0.\!''4$~pixels.  The exposure start
time is given on the image. \label{v_07}}
\end{center}
\end{figure}

The SBC is an ultraviolet imager with a field-of-view of $31'' \times
35''$ and a plate scale of $0.\!''032$~pixel$^{-1}$.  Like the ACS
CCD channels \citep[described by][]{Sirianni:2005}, the SBC focal plane
array, consisting of a multianode microchannel plate array (MAMA)
detector, is tilted with respect to the incoming rays.  This leads
to significant geometric distortion that is corrected in the calibration
pipeline processing.  The resulting asymmetric array contains the
data resampled into $0.\!''025 \times 0.\!''025$ pixels.  The pipeline
also subtracts a calibration dark count flat field image, and there is
no appreciable sky background with the F140LP filter.
A first inspection of the F140LP images reveals no concentrated source
of emission at the expected position of the comet.  It is only with
significant rebinning of the geometrically corrected images into
$16 \times 16$ superpixels (each of size $0.\!''4 \times 0.\!''4$) that
the structure of the coma emerges.  The rebinned image taken 2.69~hr
after the impact is shown in Figure~\ref{v_07}.  

\placefigure{v_07}

\subsubsection{Radial Profiles and Light Curves}

\begin{figure}
\begin{center}
\epsscale{1.0}
\rotatebox{0.}{
\plotone{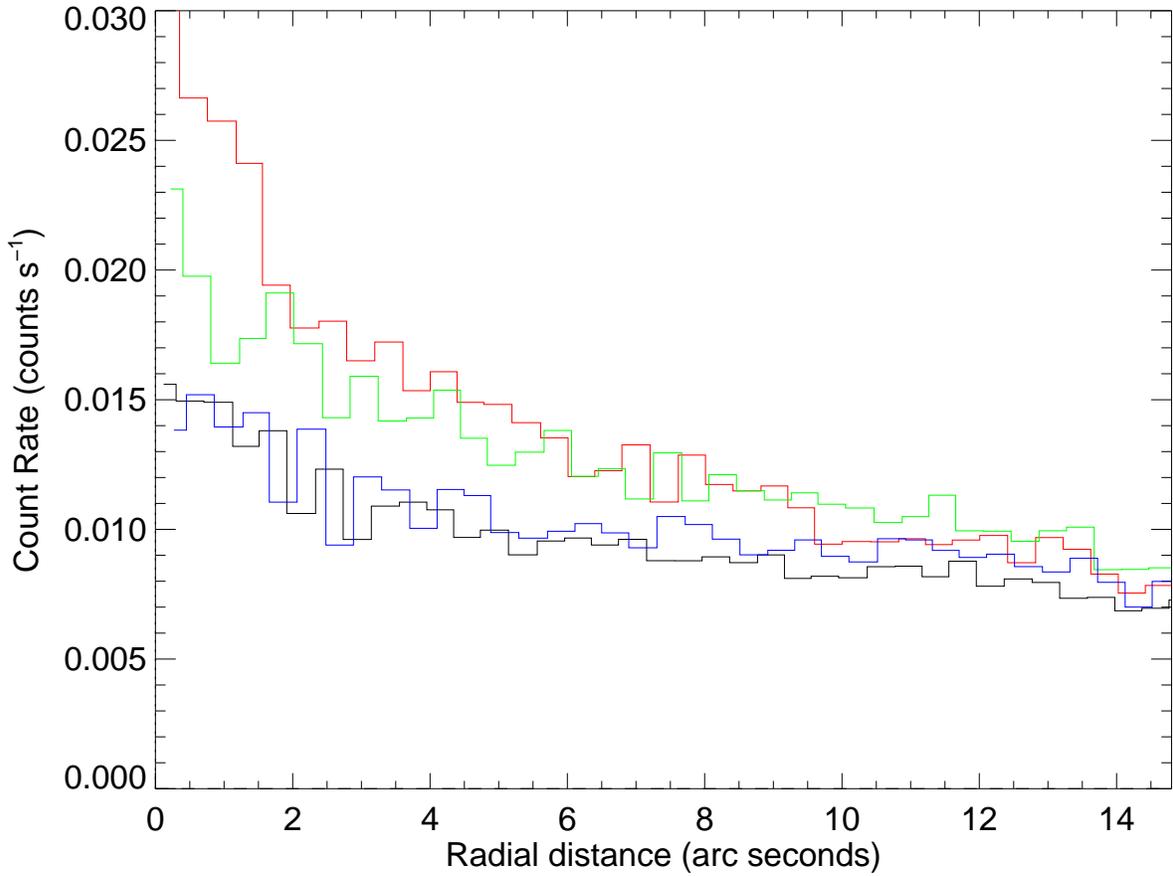}}
\vspace{-0.2in}
\caption{Radial profiles of the observed ultraviolet emission derived
from images processed as in Fig.~\ref{v_07}.  The profiles are obtained
by averaging circular annuli in widths of 1 pixel ($0.\!''4$) about the
center of brightness.  Black: average of pre-impact (visits 02 and 04)
images; red: 2.69 hr after impact; green: 5.99 hr after impact; blue:
20.29 hr after impact.  \label{rad_prof}}
\end{center}
\end{figure}

The centroid of each image is located and aperture photometry is
extracted with circular apertures of 1, 2, 4, 10 and 16 arc seconds.
Radial profiles of the emission are extracted by averaging circular
annuli of 1~pixel width over a full 2$\pi$ radians.  Directional
information is lost because of the very low count rate.  The radial
profiles are shown in Figure~\ref{rad_prof}, in which the data from the
two pre-impact images are combined to enhance the signal-to-noise
ratio of the image.

\placefigure{rad_prof}

The temporal evolution of the gas coma is illustrated by the
figure.  The visit 07 image, 2.69~hr after impact, shows an enhancement
in the inner coma with the radial profile returning to the quiescent
level $9.\!''5$ from the center of brightness.  Assuming that the CO,
like the ejected dust, is produced in a time short compared to the
time after impact \citep{A'Hearn:2005b}, then this distance
translates to a maximum component of outflow velocity of
$\sim$700~m~s$^{-1}$.  There is also evidence, in the form of a shelf
extending out $1.\!''5$ from the center of brightness, of a slower
component with a velocity of $\sim$100~m~s$^{-1}$, perhaps CO released
from slower moving small grains excavated by the impact.  However, this
only represents $\sim$10\% of the new material in the field-of-view.
The visit 09 image, 5.99~hr after impact, shows the profile decreasing
in the inner region but extending to greater distances from the nucleus
as the gas continues to expand away from the nucleus.  The final image,
from visit 11, 20.29~hr after impact, shows the gas returning to its
pre-impact level.

\begin{figure}
\begin{center}
\epsscale{1.0}
\rotatebox{0.}{
\plotone{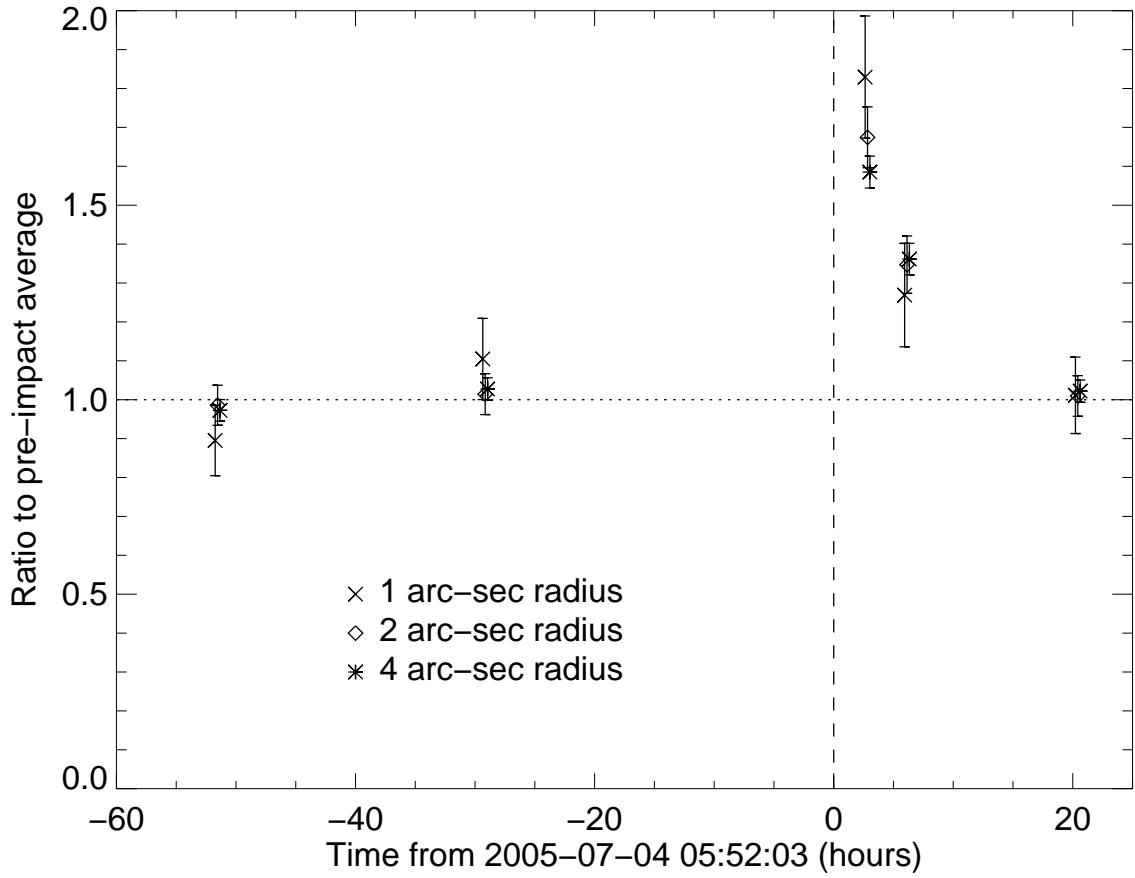}}
\vspace{-0.2in}
\caption{Light curves of the ultraviolet emission obtained by summing
the count rate in three different circular apertures.  The intensities
are plotted as the ratio to the average of the two pre-impact values.
\label{light_curve}}
\end{center}
\end{figure}

This behavior may also be illustrated with light curves derived from
the aperture photometry, as shown in Figure~\ref{light_curve}.  The
data are plotted as the ratio of the count rate to that of the
pre-impact average and for the three apertures shown the increase 
ranges from a factor of 1.8 ($1''$) to 1.5 ($4''$).  Despite the
paucity of data samples, the points for visits 07 and 09 can be fit
approximately with exponentials of time constant ranging from 3.0~hr
($1''$) to 6.5~hr ($4''$).

\placefigure{light_curve}

\subsubsection{Filter Calibration}

It is necessary to turn the observed count rate to a measure of CO
column density in the field-of-view.  From observations of many comets
by sounding rockets, \iue, and \hst, the bandpass of the F140LP filter
contains primarily emissions of \ion{C}{1} and the Fourth Positive
system (\api\ -- \xsig) of CO, with some additional weaker emissions,
dependent on heliocentric velocity, of \ion{S}{1} \citep{Feldman:2004}.  
The atomic carbon emissions, being dissociation products of long lived
species, have a flat spatial distribution near the nucleus that is
seen in \hst\ Space Telescope Imaging Spectrograph long-slit spectra
(over a $25''$ field) of several comets \citep{Weaver:2002}.  On the
other hand, the CO, except for optical depth effects close to the
nucleus, largely follows the inverse square density distribution of a
parent molecule.  For the quiescent images we assume that the spatially
varying component of the image is due solely to CO emission.

To calibrate the filter we calculate the effective transmission of the
CO Fourth Positive system as follows.  We begin with a fluorescence
model of the band system that was developed to model STIS spectra of
several comets taking into account saturation of the solar radiation
field and self absorption along the line of sight to the observer
\citep{Lupu:2004}.  For a given CO column density the calculated
line-by-line brightness of the band is multiplied by the filter
transmission at each wavelength to obtain an effective fluorescence
efficiency.  The images, which are given in counts~s$^{-1}$ per pixel,
are then converted into column density per pixel.

This process is validated using data from an ACS GTO program (ID 9985)
that observed comet C/2002~T7 (LINEAR) on 2004 June 5.  That program
included observations with two long-pass filters, F140LP and F165LP,
and with STIS, the latter to provide information about the spectral
content in the bandpass of the filters.  The difference images,
subtracting the F165LP image from F140LP, then span a wavelength range
that matches closely the long wavelength range of the STIS G140L
grating.  Using the two sets of data independently to derive a CO
production rate yields results that agree to better than 30\% of each other
(Feldman et al., in preparation).  While we cannot be certain that the
spectral content of 9P/Tempel~1 is the same as that of C/2002~T7,
particularly the content of the ejecta, within these caveats we can
reliably derive the amount of CO in the coma of 9P/Tempel~1 before and
following the impact.  Even so, it is difficult to imagine any other
species produced by the impact, such as \ntwo, that might fluoresce
strongly in the filter bandpass.

\subsection{Quiescent Production Rate}

\begin{figure}
\begin{center}
\epsscale{1.0}
\rotatebox{0.}{
\plotone{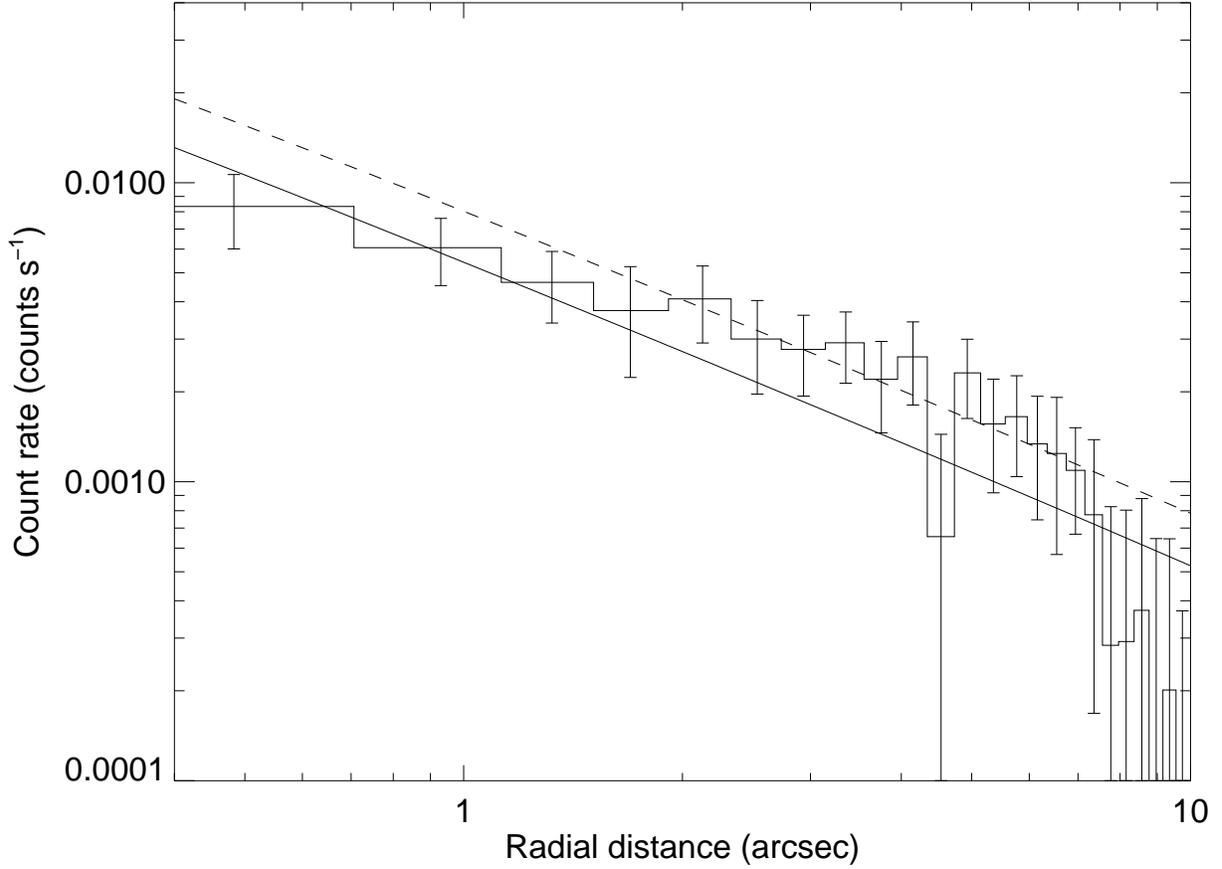}}
\vspace{-0.2in}
\caption{Pre-impact radial profile together with CO fluorescence models
that constrain the quiescent CO production rate.  A uniform background
of 0.0083 counts~s$^{-1}$ has been subtracted from the data shown
in Fig.~\ref{rad_prof}.  Solid line: Q(CO) = $4.0 \times 10^{26}$~\mols;
Dashed line: Q(CO) = $6.0 \times 10^{26}$~\mols.
\label{co_model}}
\end{center}
\end{figure}

From the mean of the two pre-impact images, a radial outflow model,
together with the effective fluorescence efficiencies described above,
is used to model the CO vaporized from a point-like nucleus.  The only
independent parameter is the outflow velocity, which we take to be the
canonically used \htwoo\ outflow velocity, $0.85 r^{-0.5}$~\kms, where
$r$ is the heliocentric distance in AU \citep{Budzien:1994}.  Even at
the edges of the approximately $30'' \times 34''$ SBC
field-of-view the observed count rate does not decrease to the detector
background level because of the atomic carbon emissions present in the
filter bandpass.  We attempt to model the radial profile with different
assumptions to constrain the level of the uniform background
emissions.  The results are shown in Figure~\ref{co_model}, and
constrain the quiescent production rate to $Q(\rm CO) = 4-6 \times
10^{26}$~\mols.  We do not consider the possibility that some of the
observed CO is the result of the dissociation of either \cotwo\ or
\htwoco.

\placefigure{co_model}

\subsection{CO Abundance in Ejecta}

The evaluation of the amount of CO in the ejecta is both easier and
more difficult than the quiescent case.  We noted that the radial
profile for visit 07 (Fig.~\ref{rad_prof}) appeared enhanced over the
pre-impact profiles only out to a radial distance of $9.\!''5$.  We can
then integrate the difference between the column densities derived for
this image and those for the pre-impact images to determine the total
CO content added to the coma.  The difficulty is the assumption that
all of the emission is due to CO, but with that assumption we derive
a total number of molecules of $1.5 \times 10^{31}$~molecules or a
mass of $6.6 \times 10^5$~kg.  This result is somewhat model dependent
although the CO column densities are for the most part too small for
unity optical depth in absorption to be reached.  Because the 
photodissociation lifetime of CO is greater than
$10^6$~s at 1~AU \citep{Huebner:1992}, the amount of atomic carbon
added to the coma by this process in $\sim$3~hr is small and does not
contribute to the observed emission.  However, we cannot exclude the
possibility of atomic sulfur produced from short-lived
parents such as H$_2$S, CS$_2$, or S$_2$, or of other ``organic''
sources of CO or C.  The infrared spectrometer on Deep Impact recorded
a large increase following impact in emission at $\sim$3.5~$\mu$m,
attributed to a C--H stretching vibration of some simple organic
molecules, and also in \cotwo\ \citep{A'Hearn:2005b}, so that some of the
observed CO could result from elevated levels of \cotwo\ or \htwoco\ in
the ejecta.  The derived abundance must thus be regarded as an upper
limit to the amount of CO produced by the excavation.  Relative
to the quiescent CO production rate given above, the added
material corresponds to 7--10~hr of natural outgassing over the
entire surface of the comet.

\section{DISCUSSION}

It is of interest to know how the CO/\htwoo\ abundance in the excavated
material compares with that found in normal cometary outgassing.  We
did not have the means to simultaneously determine the water production
rate with \hst, and the reports currently in the literature based
on different techniques are somewhat divergent.  These are summarized
in Table~\ref{tab-ratios}.  \citet{Kuppers:2005} analyzed narrow-band
OH images recorded by the OSIRIS cameras on board the {\it Rosetta}
spacecraft, and used a Haser model with a parent outflow velocity
similar to that used above for CO to obtain the quiescent production
rate.  \citet{Schleicher:2006} obtained a production rate from
ground-based OH photometry, and only a limit for the total post-impact
production.  \citet{Mumma:2005} estimated the pre-impact production
rate from infrared long-slit spectroscopy, and because their long slit
did not include all of the water produced by the impact, report only an
effective post-impact production rate which is twice their quiescent
value.  The results of \citet{Biver:2005} are from the {\it Odin}
satellite, observing at 557 GHz.

The pre-impact CO/H$_2$O value presented in Table~\ref{tab-ratios} uses
the two largest water production rates, both of which are based on
direct observations of H$_2$O.  The derived value, 4--6\%, appears to
be relatively high for a Jupiter family comet, but is well within the
range of values ($<$0.4--30\%) of several recent comets
\citep{Bockelee-Morvan:2004}.  We note that existing radio observations
do not provide a reliable constraint on the CO abundance of Jupiter 
family comets \citep{Biver:2002}. 

Combining the H$_2$O measurements post-impact with our
CO abundance yields a derived CO/H$_2$O ratio $<$10\%.  Thus, it
appears that the CO/H$_2$O abundance ratio in the excavated material is
not significantly larger than in the surface material, {\it i.e.}, the
impact did not excavate a region of buried highly volatile ices.
Our result is consistent with the findings of \citet{Sunshine:2006}
from Deep Impact images that \htwoo\ ice deposits on the surface of the
comet are insufficient to account for the ambient outgassing rate of
the comet.  They postulate extensive sub-surface sources of ice, and
it appears that the excavation of such material is also producing a brief
enhancement in CO following the impact.

We note that \citeauthor{Mumma:2005} also observed CO following the
impact (they did not measure it pre-impact), and derive a column
abundance relative to \htwoo\ in their long slit of 4.3\%, which is
consistent with our result.  However they get the same values for the
amount of CO in their slit both 1.8~hr and 24~hr after impact, a result
incompatible with our light curve (Fig.~\ref{light_curve}).
\citet{Biver:2005} also report pre- and post-impact upper limits on CO
from ground-based radio observations, but these are 3--4 times higher
than our result.

\section{CONCLUSION}

Observations of comet 9P/Tempel~1 made by the Solar Blind Channel of
the ACS on \hst\ have provided unique estimates of the CO content
of the coma both before and after the Deep Impact encounter.  Despite
the low ultraviolet flux from the comet, the observation was made
possible by the high ultraviolet throughput of the SBC and by
integration of a major fraction of the CO Fourth Positive emissions in
the bandpass of the F140LP filter.  The abundance of CO relative to
\htwoo\ remains uncertain because of the spread in the values of
\htwoo\ reported to date and the uncertain contribution of \ion{C}{1}
to the observed count rate, but to within the present uncertainties
appears to be the same in the ejecta as in the quiescent coma.  This
situation should improve upon further analysis and with additional
data.  However, it is clear that the impact did not produce ejecta
derived primarily from highly volatile ices.

\acknowledgments

We thank Ian Jordan, Ron Gilliland, and Charles Proffitt (STScI) for
the planning and successful execution of the \hst\ program; Eddie
Bergeron, Max Mutchler, Zolt Levay (STScI) and Ken Anderson (JHU) for
the rapid response and production of properly corrected images; Cheryl
Gundy, Lisa Frattare, Ray Villard, Mario Livio, and countless others at
STScI for the July 4 logistics; and STScI/JPL for coordination of
impact time before the end of \hst\ visibility.  This work was
supported by grant GO-10144.01-A from the Space Telescope Science
Institute.


\clearpage

\begin{table}
\begin{center}
\caption{Summary of ACS/Solar Blind Channel F140LP images of
comet 9P/Tempel~1. $r$ and $\Delta$ are heliocentric and geocentric
distances, respectively.  \label{co_table}}
\medskip
\begin{tabular}{@{}ccccccccc@{}}
\tableline\tableline
Visit & Image & Exposure & Date & Start & Time From &$r$ & $\Delta$ & Phase \\
& ID & Time (s) &  & Time (UT) & Impact\tablenotemark{a} (hr) & (AU) & (AU) & Angle (\degr) \\
\tableline
02 & J9A802EAQ & 509.0  & 2005-07-02  & 02:07:50 & --51.67 & 1.507 & 0.883 & 40.7 \\
04 & J9A804IGQ & 509.0  & 2005-07-03  & 00:30:59 & --29.28 & 1.506 & 0.888 & 40.8 \\
07 & J9A807F1Q & 460.0  & 2005-07-04  & 08:29:53 & 2.69 & 1.506 & 0.895 & 41.0 \\
09 & J9A809FKQ & 380.0  & 2005-07-04  & 11:48:09 & 5.99 & 1.506 & 0.895 & 41.0 \\
11 & J9A811IYQ & 509.0  & 2005-07-05  & 02:05:20 & 20.29 & 1.506 & 0.898 & 41.0 \\
\tableline
\end{tabular}
\end{center}
\tablenotetext{a}{Exposure mid-point.}
\end{table}

\clearpage

\begin{table}
\begin{center}
\caption{Pre- and post-impact \htwoo\ and CO abundances. \label{tab-ratios}}
\medskip
\begin{tabular}{@{}cccccc@{}}
\tableline\tableline
Period & H$_2$O\tablenotemark{a} & Reference\tablenotemark{b} & CO\tablenotemark{a} & Reference\tablenotemark{b}  & CO/H$_2$O \\
\tableline
Pre-impact  & $10.4 \times 10^{27}$ & 1 & $ 4-6 \times 10^{26}$ & 5 & $4-6$\% \\
       & ($8.5 \pm 1.5$)$ \times 10^{27}$ & 2 & & \\
       &  $6 \times 10^{27}$ & 3 & & \\
       & ($3.4 \pm 0.5$)$ \times 10^{27}$ & 4 & & \\
\tableline
Post-impact & ($1.4 \pm 0.35$)$ \times 10^{32}$ & 2 & $1.5 \times 10^{31}$ & 5 & $<10$\% \\
       &  $<4 \times 10^{32}$ & 3 & & \\
       &  ($1.5 \pm 0.5$)$ \times 10^{32}$ & 4 & & \\
\tableline
\end{tabular}
\end{center}
\tablenotetext{a}{Pre-impact production rates in \mols; 
post-impact abundances in molecules.}
\tablenotetext{b}{{\sc References}.--- 
(1) Mumma et al. 2005;
(2) Biver et al. 2005;
(3) Schleicher et al. 2006;
(4) Kuppers et al. 2005;
(5) This paper.}
\end{table}

\clearpage 

\clearpage 

\clearpage 

\clearpage 

\end{document}